
\input phyzzx
\def\la{\langle}
\def\ra{\rangle}
\def\uzero{{\underline 0}}

\def\uuno{{\underline 1}}
\def\udos{{\underline 2}}

\def\xz{\times}
\def\a{\alpha}
\def\b{\beta}
\def\c{\gamma}\def\C{\Gamma}
\def\d{\delta}
\def\e{\epsilon}

\def\l{\lambda}\def\Lam{\Lambda}

\def\r{\rho}
\def\s{\sigma}\def\S{\Sigma}

\def\der{\partial}
\def\uA{\underline{A}}
\def\uB{\underline{B}}

\REF\vN{P. van Nieuwenhuizen, Phys. Rev. {\bf D32} (1985) 872;\break J. Horne
and E.
Witten, Phys. Rev. Lett. {\bf 62} (1989) 501.}
\REF\AT{A. Ach\'ucarro and P.K. Townsend, Phys. Lett. {\bf 180B} (1986) 89.}
\REF\W{E. Witten, Nucl. Phys. {\bf B311} (1988) 46.}
\REF\MS{N. Marcus and J.H. Schwarz, Nucl. Phys. {\bf B228} (1983) 145.}
\REF\ATa{A. Ach\'ucarro and P.K. Townsend, Phys. Lett. {\bf B229} (1989) 383.}
\REF\Mac{S.W. Macdowell, {\it The transformation algebra of fields in
supergravity}
in proceedings of the {\sl Eighth International Conference on General
Relativity and
Gravitation}, Waterloo, Ontario, Canada, 1977.}
\REF\vNRZ{M. Ro{\v c}ek and P. van Nieuwenhuizen, Class. Quantum Grav. {\bf 3}
(1986)
43.}
\REF\NG{H. Nishino and S.J. Gates Jr., Int. J. Mod. Phys. {\bf A8} (1993)
3371.}
\REF\SW{M. Sohnius and P.C. West, Phys. Lett. {\bf 105B} (1981), 353.}
\REF\HST{P.S. Howe, K.S. Stelle and P.K. Townsend, Phys. Lett. {\bf 107B}
(1981) 82.}
\REF\DFTvN{R. D'Auria, P. Fr\' e, P.K. Townsend and P. van Nieuwenhuizen, Ann.
Phys.
(N.Y.) {\bf 155} (1984) 423.}
\REF\Wa{E. Witten, Int. J. Mod. Phys. {\bf A10} (1995) 1247.}
\REF\IT{J.M. Izquierdo and P.K. Townsend, Class. Quantum Grav. {\bf 12} (1995)
895,.}
\REF\GH{G.W. Gibbons and C.M. Hull, Phys. Lett. {\bf 109B} (1982) 190.}
\REF\Wb{E. Witten, Commun. Math. Phys. {\bf 80} (1981) 381.}
\REF\DJtH{S. Deser, R. Jackiw and G. 't Hooft, Ann. Phys. (N.Y.) {\bf 152}
(1984) 220.}
\REF\deWit{B. de Wit, H. Nicolai and A. Tollsten, Nucl. Phys. {\bf B392} (1993)
3.}
\REF\BBS{K. Becker, M. Becker and A. Strominger, {\it Three-dimensional
supergravity
and the cosmological constant}, preprint hep-th/9502107.}
\REF\bagwit{J. Bagger and E. Witten, Phys. Lett. {\bf 115B} (1982) 202.}
\REF\CG{A. Comtet and G.W. Gibbons, Nucl. Phys. {\bf B299} (1988) 719.}
\REF\SCS{B.R. Greene, A. Shapere, C. Vafa and S.T. Yau, Nucl. Phys. {\bf B337}
(1990) 1.}
\REF\HL{P.S. Howe and M.I. Leeming, {\sl in preparation}.}
\REF\OS{V. Ogievetsky and E. Sokatchev, Sov. J. Nucl. Phys. {\bf 32} (1980)
589.}
\REF\WZ{J. Wess and B. Zumino, Phys. Lett. {\bf 74B} (1978) 51.}
\REF\HLb{P.S. Howe and M.I. Leeming, Class. Quantum Grav. {\bf 11} (1994)
2843.}
\REF\KZ{D. Khetcelius and B. Zupnik, Yad Phys. {\bf 47} (1988) 1147.}
\REF\AA{V. Aldaya and J.A. de Azc\'arraga, Int. J. Theor. Phys. {\bf 24} (1985)
141.}

\font\mybb=msbm10 at 12pt
\def\bb#1{\hbox{\mybb#1}}
\def\R {\bb{R}}
\def\square{\kern1pt\vbox{\hrule height 0.5pt\hbox
{\vrule width 0.5pt\hskip 2pt
\vbox{\vskip 4pt}\hskip 2pt\vrule width 0.5pt}\hrule height
0.5pt}\kern1pt}

\font\bn=msam10 scaled 1200
\def\geq{\hbox{\bn\char '76}}
\def\leq{\hbox{\bn\char '66}}
\Pubnum{ \vbox{ \hbox{R/95/13} \hbox{hep-th/9505032}} }
\pubtype{}
\date{May, 1995, Revised December 1995}
\titlepage
\title{NEW SUPERGRAVITIES WITH CENTRAL CHARGES AND KILLING SPINORS IN 2+1
DIMENSIONS}
\author{P.S. Howe}
\address{Department of Mathematics,
King's College London,
\break
London, U.K.}
\andauthor{J.M. Izquierdo, G. Papadopoulos and P. K. Townsend}
\address{DAMTP, University of Cambridge,
\break
Silver Street, Cambridge CB3 9EW,  U.K.}
\abstract {We construct a new class of $(p,q)$-extended Poincar\'e
supergravity theories in 2+1 dimensions as Chern-Simons theories
of supersymmetry algebras with both central and automorphism
charges. The new theories have the advantage that they are limits
of corresponding $(p,q)$ adS supergravity theories and, for not too
large a value of $N=p+q$, that they have a natural formulation in
terms of off-shell superfields, in which context the distinction
between theories having the same value of $N$ but different $(p,q)$
arises because of inequivalent conformal compensator superfields.
We also show that, unlike previously constructed N-extended
Poincar\'e supergravity theories, the new (2,0) theory admits
conical spacetimes with Killing spinors. Many of our results on
(2,0) Poincar\'e supergravity continue to apply in the presence of
coupling to N=2 supersymmetric sigma-model matter.}

\endpage
\pagenumber=2
\chapter{Introduction }
It is now widely appreciated that (super)gravity theories in 2+1 dimensions can
be interpreted as Chern-Simons (CS) theories of the appropriate (super)algebra
[\vN,\AT, \W]. For example, the N-extended Poincar\'e supergravity of Marcus
and Schwarz
[\MS] can be interpreted [\ATa] as the CS theory of the standard N-extended
superalgebra
spanned by the spinor supercharges $Q^i,\, i=1\dots N$, the 3-momentum $P_a$
and angular
3-momentum $M_a$. For N=1 this algebra is the only one permitted by the
Haag-{\L}opuszanski-Sohnius theorem, but for $N\ge2$ there is the possibility
of
including central charges and it is natural to wonder whether this could lead
to new
Poincar\'e supergravity theories. Consider the $N=2$ case: in addition to the
usual
super-Poincar\'e generators one can introduce a central charge $Z$ such that
$$
\{ Q^i_\alpha,Q^j_\beta\} =-{1\over2}\delta^{ij}(\gamma^a)_{\alpha\beta} P_a +i
\epsilon_{\alpha\beta}\epsilon^{ij} Z\ .
\eqn\onea
$$
A problem with this algebra in the context of a CS formulation of N=2
Poincar\'e
supergravity is that it does not admit an invariant non-degenerate inner
product. This difficulty can be overcome as follows. We first observe that this
superalgebra has an $SO(2)$ group of automorphisms. Let $T$ be the $SO(2)$
generator;
then the only non-trivial commutator with $T$ is
$$
[T,Q^i_\alpha] = -\epsilon^{ij}Q^j_\alpha\ .
\eqn\oneb
$$
We then observe that the extension\foot{We use the word `extension' in its
technical
sense.} of the
$so(2)$ algebra by the N=2 Poincar\'e superalgebra with central charge {\sl
does} have an
invariant non-degenerate inner product; in fact, it has a three-parameter
family of such
inner products.  Let
$C$ and $A$ be the one-form gauge potentials associated to $Z$ and $T$
respectively.
Then, for any choice of the inner product, the corresponding CS
$N=2$ supergravity theory includes the new term
$$
\int CdA\ ,
\eqn\oned
$$
in addition to the usual Einstein-Hilbert and Rarita-Schwinger terms.
For reasons which will be explained shortly, we shall refer to the
Marcus-Schwarz (MS)
$N=2$ theory as the (1,1) Poincar\'e supergravity and the new $N=2$ theories as
(2,0) Poincar\'e supergravity theories. This whole discussion can be
generalized to
$N>2$ and leads to the subdivision of the $N=p+q$ Poincar\'e supergravity
theory
into the (p,q)-Poincar\'e supergravity theories. We present this more general
construction in section 7.
One motivation for the introduction of central charges in the Poincar\'e
superalgebra
comes from consideration of the (p,q) anti de Sitter (adS) supergravity
theories,
which include an $SO(p)\times SO(q)$ CS term and a cosmological constant
proportional
to $m^2$, where $m$ is a gravitino `mass' parameter. The (p,q)-extended
Poincar\'e
superalgebra is found in the limit as $m\rightarrow 0$. Even so, the
$m\rightarrow 0$
limit of the (p,q) adS supergravity {\sl action} of [\AT] is problematic when
either $p$
or $q$ is greater than one; the only non-singular way to take the $m\rightarrow
0$ limit
causes the $SO(p)\times SO(q)$ gauge fields to disappear from the action (which
is then
the N=p+q extended MS Poincar\'e supergravity) but leaves a non-zero
supersymmetry
transformation for them [\ATa]. This means that the incorporation of a
cosmological
constant into the $N>2$ MS theories cannot be done without the introduction of
additional fields, which is a state of affairs that could never arise from
elimination
of auxiliary fields in an off-shell supergravity theory. One purpose of this
paper
is to show how this difficulty is resolved by consideration of a trivial
extension of
the (p,q) adS superalgebra by an $so(p)\oplus so(q)$ algebra. A particular
$m\rightarrow 0$
limit of these new (p,q) adS supergravity theories yields the (p,q) Poincar\'e
ones.
Thus the (p,q) Poincar\'e supergravity theories introduced here naturally arise
as
limits of (p,q) adS supergravity theories.
Another motivation for the new CS formulation of Poincar\'e and adS
supergravities
comes from consideration of off-shell supersymmetry. Since it is obvious that
the
algebra of CS gauge transformations already closes off-shell, i.e. without the
use
of the field equations, and that these transformations include spinorial ones,
it
might be thought that there is nothing to be considered here. However, it is
easy to
verify that the numbers of off-shell boson and fermion fields (subtracting the
gauge
degrees of freedom) are not generally equal, which is puzzling since this is
usually
thought to be a precondition of off-shell supersymmetry. The resolution of this
puzzle lies in the correspondence between the gauge symmetries of the CS theory
and
the geometrical symmetries of the supergravity theory, which contain spacetime
diffeomorphisms. This correspondence is an {\it on-shell} one [\ATa], so that
off-shell closure of the CS gauge transformations does not imply off-shell
closure of
the geometrical supersymmetry transformations. It is the latter that is needed
for a superspace formulation, and this requires equality of off-shell boson and
fermion degrees of freedom. This shows, incidentally, that there are, in
principle,
{\it two} ways in which off-shell closure of the algebra of local supersymmetry
transformations may be achieved in theories for which given supersymmetry
transformations form a closed algebra only on-shell. One can try to close the
algebra
by the addition of `trivial' transformations or one can try to close it by the
addition of auxiliary fields, which is tantamount to a reformulation in
terms of superfields. The former method was actually advocated in an early
attempt to
obtain a closed off-shell algebra for $N=1$ supergravity in 3+1 dimensions
[\Mac] but
the attempt was unsuccessful because in spacetimes of dimension 3+1, or
higher, the
addition of auxiliary fields is unavoidable. In 2+1 dimensions one can always
close
the algebra in {\it pure} supergravity theories by adding `trivial' gauge
transformations and this amounts to a reformulation as a CS theory, but this is
unlikely to work once matter is included. We shall consider the supersymmetry
algebra
to be closed off-shell only if the geometrical supersymmetry transformations
form a
closed algebra, which effectively means that we consider a given theory to be
off-shell supersymmetric only if it has an off-shell superfield formulation. It
is by
no means guaranteed that such a formulation exists; considerable difficulties
appear
for (p,q) theories when either $p$ or $q$ exceeds 2, as we shall see.
To explain why off-shell supersymmetry is relevant to (p,q) Poincar\'e
supergravity it is simplest to consider the N=2 case, which subdivides into the
(2,0) and the (1,1) cases. The possibility of two inequivalent theories arises
in the
N=2 superfield context from the fact that the conformal invariance of the
conformal
supergravity supermultiplet can be `compensated', to produce a Poincar\'e
supermultiplet, by two inequivalent `compensating superfields'. The N=2
conformal
supergravity supermultiplet has the field [dimension] content [\vNRZ]:
$$
e_m{}^a\, [0]\ ;\ \psi^i_m\, [{1\over2}]\  ; \ A_m\, [1]\ ,
\eqn\onee
$$
where $e_m{}^a$ is the dreibein, $\psi^i_m$ the two gravitini, $A_m$ the SO(2)
gauge
potential and the dimension is the `geometrical' one. One possible N=2
compensator
multiplet has the field [dimension] content
$$
A\, [0]\, ,\, B\, [0] \ ;\ \lambda^i\, [{1\over2}]\ ;\ \tilde K_{ij}\, [1] \ ,
\eqn\onef
$$
where $\tilde K_{ij}$ is symmetric and traceless in its two SO(2) indices. The
scalars
$A$  and $B$ become the trace of the dreibein and the longitudinal component of
$A_m$,
respectively, while $\lambda^i$ becomes the gamma-trace of $\psi_m^i$. This
yields the
Poincar\'e supermultiplet with the following field [dimension] content
$$
e_m{}^a\, [0]\ ;\ \psi^i_m\, [{1\over2}]\  ; \ A_m\, [1]\, ,\, \tilde K_{ij}\,
[1]
\eqn\oneg
$$
where $A_m$ and $\tilde K_{ij}$ are both auxiliary fields. The off-shell
Poincar\'e and
adS supergravity theories constructed from this supermultiplet are those of
(1,1)
supersymmetry [\NG].
Another possible compensating supermultiplet has the field [dimension] content
$$
A\, [0]\, ,\,  C_m\ [0]\ ;\ \lambda^i\, [{1\over2}]\ ; K\, [1]\ .
\eqn\oneh
$$
The scalar $A$ and spinors $\lambda^i$ get absorbed as before, so that the
field
[dimension] content of the resulting Poincar\'e supermultiplet is
$$
e_m{}^a\, [0]\, , C_m\, [0]\  ;\ \psi^i_m\, [{1\over2}]\  ; \ A_m\, [1]\, ,\,
K\, [1]\ ,
\eqn\onei
$$
where the scalar $K$ is auxiliary but the vectors $A_m$ and $C_m$ are both
gauge fields
which can be identified as the `Automorphism' and `Central' gauge fields of the
(2,0)
Poincar\'e supergravity, i.e. the components of the gauge potential one forms
$A$ and
$C$ introduced above. Indeed, one finds that the off-shell supersymmetric
action
contains precisely the term \oned, as expected for the (2,0) theory.
There is a precedent for such a term in the new-minimal formulation of
off-shell N=1
supergravity in 3+1 dimensions, where $C_m$ is replaced by a two-form gauge
potential
[\SW,\HST,\DFTvN]. In that case, it is customary to consider the $C$ and $A$
gauge fields
as auxiliary, despite the fact that their field equations are not algebraic,
because
on-shell they are pure gauge. While this is justifiable in 3+1 dimensions it is
obviously
not in 2+1 dimensions because, by the same criterion, all CS gauge fields,
including the
dreibein, would have to be considered as auxiliary. We may conclude from this
that in 2+1
dimensions different choices of the conformal compensator supermultiplets can
lead to
{\sl inequivalent} Poincar\'e theories.
As for all supergravity theories, the gravitino supersymmetry transformation of
an
N=2 Poincar\'e theory is a covariant derivative of the supersymmetry parameter
$\zeta$. In complex spinor notation we can write it as $\delta_\zeta\psi =
{\cal D}\zeta$ where ${\cal D}$ is a covariant derivative acting on complex
spinors.
For the (1,1) Poincar\'e theory, and in a purely bosonic background, ${\cal D}$
is just
the standard covariant derivative constructed from the spin connection. One can
define a
supersymmetric spacetime to be one for which all fermions {\sl and their
supersymmetry variations} vanish for some non-zero supersymmetry parameter. A
necessary condition is therefore that there exist a nonzero spinor $\zeta$ such
that
${\cal D}\zeta=0$. Since this equation is linear in $\zeta$, its consequences
are unchanged if we replace the anticommuting spinor $\zeta$ by a commuting one
$\kappa$. Such a spinor, satisfying ${\cal D}\kappa=0$, is generally called a
Killing
spinor. In the context of (1,1) supergravity a Killing spinor is one that is
covariantly constant with respect to the usual spin-connection, which is
possible only
if the spacetime is flat. Solutions of the vacuum field equations of all
Poincar\'e
supergravity theories outside a matter source are pure gauge configurations on
conical
spacetimes with a deficit angle equal to the mass $M$ in the interior; we
therefore
restrict $M$ to lie in the range $0\leq M<2\pi$. Since these spacetimes are
flat one
might expect them to be supersymmetric but, as recently observed [\Wa], there
are no
covariantly constant spinors in conical spacetimes unless $M=0$, so the only
supersymmetric spacetime in the context of the (1,1) Poincar\'e supergravity is
2+1
Minkowski spacetime, with $M=0$, and, evidently, the same conclusion holds for
the
N-extended MS supergravity theories for any $N$. Another purpose of this paper
is to
point out that this conclusion changes when one considers the new (2,0)
Poincar\'e
supergravity. Here the covariant derivative ${\cal D}$ includes the
automorphism
gauge field $A$. The corresponding charge, $Q$, can be chosen, without loss of
generality, to satisfy $|Q|<\pi$. It is now possible to find spinors that are
covariantly constant outside the region containing the matter provided that
$$
{\it either}\qquad M=2|Q| \qquad {\it or}\qquad M= 2|Q\pm\pi|\ ,
\eqn\onej
$$
depending on whether the spin structure is even or odd, respectively. In
particular,
conical point-particle spacetimes admit Killing spinors for any allowed value
of $M$,
and multi point-particle spacetimes may do so also, provided a condition
analogous to
\onej\ holds for every particle.
Matter coupling to N-extended supergravity theories with zero cosmological
constant
has been considered in [\deWit] and with non-zero cosmological constant for N=2
in [\IT].
More recently, an N=2 matter coupled three-dimensional supergravity was
constructed by dimensional reduction from four dimensions [\BBS], and the
possibility of
Killing spinors in this context was demonstrated. Here we consider the coupling
of (2,0)
Poincar\'e supergravity to sigma-model matter. An interesting feature of this
model is
that while the automorphism gauge field $A$ couples to the $U(1)$ Noether
current, the
central charge gauge field $C$ couples to a topological current.
In supergravity theories one generally finds that supersymmetric spacetimes,
i.e.
those admitting Killing spinors, saturate a bound on the mass in terms of the
charges. Of course, supergravity theories in 2+1 dimensions are exceptional in
many
respects, but such a bound has recently been established for adS supergravity
theories in 2+1 dimensions [\IT]. This might lead one to suppose that \onej\
has an
interpretation as the saturation by supersymmetric spacetimes of a lower {\it
bound}
on the mass $M$ of any spacetime solving the equations of a matter-coupled
(2,0)
Poincar\'e supergravity, for a given choice of spin structure for the
gravitino. The
standard way of deriving a bound on the mass in terms of the charges in
supergravity
theories proceeds via a generalization [\GH] of the spinorial proof of the
positive
energy theorem. A curious feature of the attempt, which we  describe later, to
obtain
such a bound in the present case is that the `Witten condition' that must be
imposed
on the spinor appearing in the expression for the total energy [\Wb] turns out
to be
equivalent to the condition that this spinor be a Killing spinor. Thus, instead
of
deriving a bound one merely confirms the equality \onej.
The organisation of this paper is as follows: In section 2 we describe the CS
formulation of the (2,0) Poincar\'e supergravity theories. In section 3 we
discuss
Killing spinors of the (2,0) Poincar\'e supergravity theory. In section 4 we
couple the
(2,0) supergravity theory to sigma-model matter. In section 5 we discuss the
application
of the Witten-Nester method to 2+1 gravity, and prove an energy `equality'
theorem.
In section 6  we explain the connection with off-shell superfields and present
the
superfield construction of (2,0) Poincar\'e supergravity coupled to the most
general
sigma-model matter. In sections 7 and 8  we present the CS formulation of the
general (p,q) Poincar\'e supergravity and discuss their relation to (p,q) adS
supergravity.
\chapter{CS formulation of (2,0) Poincar\'e Supergravity}
The non-trivial commutators of the (2,0) Poincar\'e superalgebra and its
$so(2)$
automorphism algebra are
$$
\eqalign{
[M_a,M_b] &=-\epsilon_{ab}{}^cM_c\cr
[M_a,P_b] &=-\epsilon_{ab}{}^cP_c\cr
\{ Q^i_\alpha,Q^j_\beta\} &=-{1\over2}\delta^{ij}(\gamma^a)_{\alpha\beta} P_a
+i
\epsilon_{\alpha\beta}\epsilon^{ij} Z\cr
[T,Q^i_\alpha]& = -\epsilon^{ij}Q^j_\alpha\ .}
\eqn\twoa
$$
We use the `mostly-minus' metric convention and hence gamma matrices that are
pure imaginary. Note that
$$
\gamma^a\gamma^b =\eta^{ab} + i\epsilon^{abc}\gamma_c\ \qquad
(\bar\psi)_\alpha=\psi^\beta \epsilon_{\beta\alpha}.
\eqn\twob
$$
We also introduce a formal conjugation with respect to which all the even
generators
of the superalgebra are antihermitian whereas the odd ones are hermitian, and
we adopt the
standard convention that complex conjugation of a fermion bi-linear introduces
an
additional minus sign.  Defining the (antihermitian) connection\foot{Note that
in our
conventions the field
$\psi$ anticommutes with $Q$.} one-form
$a$ by
$$
a= e^aP_a + \omega^aM_a + CZ +AT + \psi^iQ_i\ ,
\eqn\twoc
$$
we compute the curvature two-form $F=da+a^2$ to be
$$
F= T^aP_a + F^a(\omega)M_a +F(C)Z + F(A) T + {\cal D}\psi^i Q_i\ ,
\eqn\twod
$$
where
$$
\eqalign{
T^a &= de^a - \epsilon^a{}_{bc} \omega^b e^c -{1\over4}\psi^i\gamma^a\psi^i \cr
F^a(\omega) &=d\omega^a -{1\over2}\epsilon^a{}_{bc}\omega^b\omega^c\cr
F(C) &=dC\cr
F(A) &=dA \cr
{\cal D}\psi^i  &= d\psi^i +{i\over2} \omega^c\gamma_c\psi^i +
A\epsilon^{ij}\psi^j\ . }
\eqn\twoee
$$
There is a three-parameter non-degenerate inner product on the above algebra,
which is
given as a special case of the four-parameter inner product for the general
N-extended case
in section 5. The freedom represented by the choice of these parameters is not
significant for the classical theory considered here, so we shall make the
`simplest'
choice for which the only non-zero components of the inner product, in the
above basis,
are
$$
\la M_a,P_b\ra = \mu\, \eta_{ab} \qquad \la Q^i_\alpha,Q^j_\beta\ra =i \mu\,
\epsilon_{\alpha\beta}\delta^{ij}\qquad \la T,Z\ra =-\mu\ ,
\eqn\twoc
$$
where $\mu$ is a real non-zero constant with dimensions of mass and the inner
product is
hermitian with respect to the formal conjugation introduced above. The CS
action is then
$$
S=\mu\int d^3x\ \big[  eR -i  \varepsilon^{mnp}\bar\psi^i_m{\cal
D}_n\psi^i_p +2  \varepsilon^{mnp}C_m\partial_n A_p \big]\ ,
\eqn\twod
$$
where we consider the spin-connection $\omega_m{}^a$ to be an independent
field, i.e.
this is the first-order form of the supergravity action.
This action is invariant under the gauge transformations of the connection $a$
and in particular under local supersymmetry transformations.  The non-zero
supersymmetry transformations are
$$
\eqalign{
\delta e^a &= {1\over 2}\bar\zeta^i\gamma^a \psi^i\cr
\delta \psi^i &= {\cal D}\zeta^i\cr
\delta C &= i\epsilon_{ij}\bar\zeta^i\psi^j\ , }
\eqn\twoe
$$
where $\zeta^i$ are anticommuting spinor parameters.
The field equations of  \twod\ are equivalent to the vanishing of the
components,
\twoee, of the curvature two-form $F(a)$. We shall be interested in solutions
of these (vacuum) equations for which the spacetime is asymptotically conical
at
spatial infinity, i.e. the  metric is asymptotic to
$$
ds^2=dt^2-\lambda^2 dr^2-r^2
d\varphi^2\quad,\quad\lambda>1 \ ,
\eqn\twof
$$
as $r\rightarrow \infty$. From the CS formulation of the theory, one can deduce
[\IT]
that the total mass relative to Minkowski spacetime is given
by\foot{Underlining
indicates a {\it frame} index, i.e. $a=({\underline 0},{\underline i})$,
whereas
$m=(0,i)$.}
$$
M=-\oint_\infty (\omega^\uzero-\bar\omega^\uzero) \ ,
\eqn\twog
$$
where $\bar\omega$ is the Minkowski spacetime spin connection. Note that the
term
involving $\bar\omega$ depends on the choice of frame and is not necessarily
zero, as can
easily be verified by choosing the natural frame of polar coordinates. From
this
definition one may verify the result of [\DJtH] that the mass of an
asymptotically-conical
spacetime equals the deficit angle, i.e.
$$
M=2\pi \big( 1-{1\over \lambda}\big)\ .
\eqn\twoh
$$
We also require that the vector field $A$ be asymptotic to $A_\varphi^\infty
d\varphi$,
where $A_\varphi^\infty$ is a constant. The total $A$-charge can be defined by
$$
Q= \oint_\infty A = 2\pi A_\varphi^\infty\ .
\eqn\twoi
$$
In the case that the circle at infinity is contractible, all $U(1)$ gauge
transformations, $A\rightarrow A'= A+d\alpha$, with $d\alpha ={\cal O}(1/r)$,
are
connected to the identity. In this case the charge $Q$ is clearly
gauge-invariant. If,
however, the circle at infinity is not contractible, as occurs for conical
point-particle
spacetimes, then there are additional `large' gauge transformations, i.e. those
not
connected to the identity, and the charge $Q$ is not gauge invariant but can
change by a
multiple of $2\pi$. In these circumstances, the charge $Q$ is well-defined only
modulo
$2\pi$.
\chapter{Killing Spinors}
A bosonic configuration of (2,0) Poincar\'e supergravity  will preserve some of
the
supersymmetry if there is a non-zero $\zeta$ such that $\delta_\zeta\psi^i=0$.
Since
this equation is linear in the anticommuting spinor parameters $\zeta^i$ its
consequences are unchanged if $\zeta^i$ is replaced by commuting spinors. It is
also
convenient to choose a complex basis for these commuting spinors. In such a
basis,
`supersymmetric' bosonic configurations are those backgrounds for which the
equation
$$
(D - i A)\kappa=0
\eqn\threea
$$
has non-trivial solutions for complex commuting spinor $\kappa$. The
integrability condition for this equation is
$$
\big[F^a{}_{mn}(\omega) \gamma_a -2 F_{mn}(A)\big]\kappa =0\ .
\eqn\threeb
$$
This integrability condition is satisfied by all configurations  that satisfy
the vacuum
equations.  Although \threeb\ is a necessary condition for the existence of
solutions of
\threea\ it is not sufficient; additional conditions must be satisfied in order
for
\threea\ to have non-trivial solutions for $\kappa$. To find these additional
conditions,
we first write the general solution of \threea\ for $\kappa$ at a point $p$, in
terms of
its value at a point $q$, as follows:
$$
\kappa(p) = {\cal P}exp\Big[\int_C \big(-i{1\over2}\omega^a\gamma_a +
iA\big)\Big]\kappa(q)
\eqn\threec
$$
where $C$ is a path that
interpolates between $q$ and $p$, and ${\cal P}$ indicates path-ordering of the
exponential. This solution is not necessarily single-valued  because it may
depend on the path $C$.  However, since the connections $\omega$ and $A$ are
flat the
solution can depend only on the homotopy class, $[C]$, of the path $C$. To
compute the
path dependence of the solution for $\kappa$, let $C'$ be another path between
$q$ and
$p$. Then $\kappa$ is well-defined if and only if the condition
$$
\lambda\big([\Gamma]\big)\kappa(q)
={\cal P}exp\Big[\oint_\Gamma \big(-i{1\over2}\omega^a\gamma_a +
iA\big)\Big]\kappa(q)
\eqn\threed
$$
is satisfied, where $\Gamma= (C')^{-1}\cdot C$ is a closed path from $q$ to $q$
and
$\lambda\big([\Gamma]\big)$ is a sign depending on the choice of spin
structure. The choice of spin structure corresponds to the choice of periodic
or
anti-periodic boundary conditions for the spinors on the non-trivial
fundamental homology
cycles of space-time. The spin structure corresponding to periodic boundary
conditions is
called `even', and the spin structure corresponding to antiperiodic boundary
conditions is
called `odd'.  The condition \threed\ is trivially satisfied if the spacetime
is simply
connected;  the holonomies of both $A$ and $\omega$ are then equal to one. The
interesting case is when the spacetime is not simply connected, in which case
the
existence of covariantly constant spinors depends on the holonomy of the flat
connections
$A$ and $\omega$ and also on the choice of spin structure. If only one of the
connections
$\omega$ and $A$ has non-trivial holonomy the equation \threed\ does not have
solutions,
and the  vacuum solution will not be supersymmetric, so it remains to examine
the case
that both connections have non-trivial holonomy. The holonomy of $\omega$
evaluated at
any closed path is an element of $SL(2,\R)$.  This holonomy is called
elliptic,
parabolic or hyperbolic depending on whether the value of the trace of its
holonomy
$2\times 2$ matrix is less than, equal to, or larger than two, respectively. It
is clear
that in order to satisfy \threed\ the holonomy of $\omega$ must be elliptic.
So, possibly
after a gauge transformation, \threed\ can be rewritten as
$$
\lambda\big([\Gamma]\big) \kappa(q)
={\cal P}exp\Big[\oint_\Gamma \big(-i{1\over2}\omega^{\underline
0}\gamma_{\underline 0} + iA\big)\Big]\kappa(q)\ .
\eqn\threee
$$
Next, we pick $\kappa(q)$ to satisfy
$$
\gamma_{\underline 0}\kappa(q)=\pm \kappa(q)  \ ,
\eqn\threef
$$
in which case the equation \threee\ becomes
$$
\lambda\big([\Gamma]\big)\kappa(q)
={\cal P}exp\Big[\oint_\Gamma \big(\mp i{1\over2}\omega^{\underline 0} +
iA\big)\Big]\kappa(q)\ .
\eqn\threeg
$$
which is satisfied if and only if the holonomy of $\mp
{1\over2}\omega^{\underline 0}$
cancels the holonomy of $A$ up to a sign, depending on the choice of even or
odd spin
structure. We will refer to this as the `holonomy condition'.
In 2+1 dimensions, the static multi point-particle spacetime is flat everywhere
except at
isolated conical singularities.  These spacetimes are non-simply connected if
one removes
the singular worldlines of the $K$ particles; space is then topologically a
plane with $K$
punctures. The metric of such a spacetime is
$$
ds^2= dt^2-h^2 \delta_{ij}dx^i dx^j\ , \qquad i,j=1,2\ ,
\eqn\threeh
$$
where
$$
h^2=\prod^K_{\ell=1} |x-x_\ell|^{-{m_\ell\over \pi}}\ ,
\eqn\threei
$$
$m_\ell$ are the masses of the particles and $r_\ell$ are their positions.
The only non-zero component of the connection $\omega$ is
$$
\omega^{\underline 0}_i=\epsilon^{jk}\delta_{ki} h^{-1}\partial_j h\ .
\eqn\threej
$$
It can be shown using Cauchy's theorem that the holonomy of
$\mp{1\over2}\omega^{\underline 0}$ evaluated on a path $\Gamma_n$ that
encloses $n\le
K$ particles with masses
$m_1,\dots, m_n$ is
$$
exp\Big[{\pm {i\over2} M_n }\Big]\ ,
\eqn\threek
$$
where $M_n=\sum_{\ell=1}^n m_\ell$.  There are $2^K$ different spin structures
on this
spacetime. If we define the charge $Q_n$ of the $n$ point particles by
$$
Q_n=\oint_{\Gamma_n}\! A\ ,
\eqn\threel
$$
then the holonomy condition implies that
$$
M_n=\mp 2 Q_n+2\epsilon([\Gamma_n])\pi \nu
\eqn\threem
$$
where $\nu$ is an integer and $\epsilon([\Gamma_n])$ is equal either to $1$ or
to $2$,
depending on whether the spin structure is chosen to be odd or even,
respectively. Recalling that $Q_n$ is defined only mod $2\pi$, and that $M=M_K$
is
restricted by $0\leq M< 2\pi$ , we see that for an even spin structure \threem\
is
equivalent to
$$
M_n= 2|Q_n|\ ,
\eqn\threen
$$
where we choose $Q_n$ such that $|Q_n| < \pi$. For an odd spin structure we can
choose $\nu$ such that
$$
M_n= 2|Q_n \pm \pi|\ ,
\eqn\threeo
$$
where the sign must be chosen to ensure that $M<2\pi$. In either case the
relevant
holonomy condition must hold for every closed path, and this implies that the
mass of
each particle is related to its charge as above. In particular, we find that
the total
mass and charge are related as in \onej. Thus, {\it the conditions on the
masses and
charges required for the existence of a Killing spinor depend on the choice of
spin
structure}. We remark that the case of one particle with mass $M=\pi$ and
charge $Q=\mp{1\over2}\pi$ is special in that the corresponding spacetime
admits
Killing spinors of either spin structure.
We conclude this section with a remark. First, note that a spinor that is
Killing
in some region of spacetime, e.g. an asymptotic region, cannot necessarily be
extended
to a Killing spinor of the entire spacetime. This is because the holonomy
matching
condition should hold for {\it every} fundamental homology cycle of spacetime.
\chapter{Matter coupling to (2,0) supergravity}
We shall now consider the coupling of the general N=2 sigma model to (2,0)
Poincar\'e
supergravity. As explained in the introduction, the off-shell supermultiplet
includes the
gauge field one-forms $A$ and $C$ and an auxiliary scalar field $K$ in addition
to the
dreibein one-form $e^a$ and the two Majorana gravitini, which we may exchange
for a
single Dirac spinor one-form $\psi$. The Lagrangian is
$$
{\cal L}_{grav}= -eR +i \varepsilon^{mnp}\bar\psi_m \hat{\cal
D}_n\psi_p -2 \varepsilon^{mnp}C_m\partial_n A_p + 2e K^2
\eqn\mone
$$
where
$$
\hat{\cal D}_m = {\cal D}_m - {1\over 4}e^{-1} \varepsilon_{mnp}\gamma^n
V^p + {i\over2}\gamma_m K
\eqn\mtwo
$$
and the Dirac conjugate is defined by $\bar\psi =\psi^{\dagger}
(i\gamma^{\underline 0})$,
which is equivalent to the previous definition if $\psi$ is real. This
Lagrangian is
invariant, up to a total derivative, under the following supersymmetry
transformations
$$
\eqalign{
\delta e_m{}^a &= {1\over4}(\bar\epsilon\gamma^a\psi_m -
\bar\psi_m\gamma^a\epsilon )\cr
\delta \psi_m &= \hat{\cal D}_m\epsilon \cr
\delta C_m &= -{1\over2}\bar\psi_m\epsilon + {1\over2}\bar\epsilon\psi_m\cr
\delta A_m &= {i\over 8} e^{-1}\varepsilon_{mnp}\epsilon\gamma^n\chi^p + c.c.
\cr
\delta K &= -{1\over8}\bar\epsilon\gamma\cdot\chi + c.c. }
\eqn\mthree
$$
where
$$
\eqalign{
V^m &\equiv e^{-1}\varepsilon^{mnp}\partial_n C_p \cr
\chi^m &\equiv e^{-1}\varepsilon^{mnp} {\cal D}_n\psi_p }
\eqn\mfour
$$
Note that the Lagrangian \mone\ is equivalent (for $\mu=-1$) to that of \twod\
after
elimination of the auxiliary field $K$ by its algebraic equation of motion.
Note also
that the supersymmetry transformations \mthree\ differ from the CS
supersymmetry
transformations by the addition of transformations proportional to the field
equations.
The transformations given here are the `geometrical' ones. Since these
transformations
close to give general coordinate and other gauge tranformations without the use
of the
field equations, the coupling to matter will not change them.
The three-dimensional off-shell N=2 scalar multiplet has the field content
$(\Phi,\lambda, F)$ where $\Phi$ is a physical complex scalar, $\lambda$ a
Dirac spinor
and $F$ a complex auxiliary field. The coupling of (2,0) supergravity to n such
multiplets
leads to a (2,0) locally supersymmetric sigma-model with an $n$
complex-dimensional
K\"ahler target space. One method of constructing this model would be through
the
development of a tensor calculus, which is effectively equivalent to the
superfield method developed in section 6.  Here we shall make use of the
order-by-order `Noether method' ; since the matter auxiliary fields play no
role in this
method they will be omitted from the start. It is also convenient to introduce
$2n$ real
scalar fields
$\phi^I$, instead of
$n$ complex scalar fields. Thus, the matter fields that we shall need to
consider
are
$$
(\phi^I,\ I=1\dots 2n\ ;\ \lambda^A,\bar\lambda_A, \ A=1\dots n)
\eqn\mfive
$$
where
$$
\bar\lambda_A = (\lambda^A)^{\dagger}(i\gamma^{\underline 0})\ .
\eqn\msix
$$
With this notation it is necessary to introduce a target space vielbein
$\Big(f_I{}^A,
f_{IA}= \overline{(f_I{}^A)}\Big)$ and its inverse $\Big(f^I{}_A, f^{IA}=
\overline{(f^I{}_A)}\Big)$ satisfying
$$
\eqalign{
&f^I{}_A f_I{}^B = \delta_A^B \qquad\qquad f^{IA}f_{IB}= \delta^A_B\cr
&f^I{}_A f_I{}_B = 0 \qquad\qquad f^{IA}f_I{}^B=0 \cr
& f^I{}_Af_J{}^A + f^{IA}f_{JA} = \delta^I_J \ .}
\eqn\mseven
$$
The K\"ahler two-form $\Omega$ can be defined via the relation
$$
g_{IJ} -i\Omega_{IJ} = 2f_I{}^A f_{JA}
\eqn\meight
$$
where $e=\det e_m{}^a$ and $g$ is the target space metric. We refer to [\IT]
for more
details of the notation.
The locally (2,0)-supersymmetric sigma-model Lagrangian for fields \mfive\ is
$$
\eqalign{
{\cal L}_{\sigma} = {1\over2}& e\;\big[ g_{IJ}\partial_m\phi^I\partial^m\phi^J
-
\bar\lambda_A\gamma^m\nabla_m\lambda^A \big] + {1\over2}C_mj^m_{top}\cr
&-{i\over2} e\; \big[\bar\lambda_A\gamma^m\gamma^n\psi_m f_I{}^A +
\bar\psi_m\gamma^n\gamma^m\lambda^A f_{IA}\big]\partial_n\phi^I
+ ({\rm quartic\ fermion\ terms})}
\eqn\mnine
$$
where $\nabla = {\cal D} + L$, with $L$ equal to the spin-connection of the
Levi-Civita connection with respect to the frame $f$. Notice that the central
charge
gauge potential $C$ couples to the topological current
$$
j^m_{top} \equiv {1\over2} \varepsilon^{mnp}\Omega_{IJ}\partial_n\phi^I
\partial_p\phi^J\ .
\eqn\mten
$$
We have not made any attempt to calculate the quartic fermion terms because
they are not
important for our present purposes. The complete construction, using superfield
methods,
will be presented in section 6. The action
\mnine\ is invariant under the transformations of the supergravity fields given
above
together with the following supersymmetry transformations of the matter fields:
$$
\eqalign{
\delta\phi^I &= {i\over2}f^I{}_A\bar\epsilon\lambda^A + {i\over2} f^{IA}
\bar\lambda_A\epsilon \cr
\delta\lambda^A &= -{i\over2} f_I{}^A\gamma^m\epsilon\; \partial_m\phi^I +
({\rm cubic\
fermion\ terms}) \ .}
\eqn\meleven
$$
The Lagrangian for the coupling of (2,0) Poincar\'e supergravity to N=2
sigma-model matter
is simply the sum ${\cal L} = {\cal L}_{grav} + {\cal L}_{sigma}$ from which
the
supergravity auxiliary field $K$ may be trivially eliminated (neglecting
possible
additional quartic fermion terms) to arrive at the Lagrangian (in a condensed
notation)
$$
\eqalign{
{\cal L} = e\Big[ -R +  {1\over2} (\partial\phi)^2
-\bar\lambda\gamma\cdot\nabla\lambda
-\big\{{i\over2}\partial_n\phi\cdot \bar\lambda \gamma^m\gamma^n\psi_m +
c.c.\big\}\Big]
\cr +i\varepsilon^{mnp}\bar\psi_m{\cal D}_n\psi_p
-2\varepsilon^{mnp}C_m\partial_n A_p +
{1\over2} C_mj^m_{top} \ .}
\eqn\mtwelve
$$
For what follows we shall need only the bosonic field equations of this
Lagrangian, and
the fermion transformation laws, in backgrounds for which the fermion fields
vanish. The
bosonic field equations are
$$
\eqalign{
\varepsilon^{mnp}\partial_nA_p &= {1\over4} j^m_{top}\cr
G_{mn} &= {1\over2} T_{mn}\cr
V^m &=0\cr
\nabla^2\phi &=0 \ .}
\eqn\mthirteen
$$
where $T_{mn}$ is the energy-momentum tensor
$$
T_{mn} = \partial_m\phi\cdot \partial_n\phi - {1\over2}g_{mn}\partial\phi\cdot
\partial\phi\ .
\eqn\mfourteen
$$
The first of these equations states that the two-form field strength, $F=dA$,
is proportional to the pullback of the K\"ahler two-form, $\Omega$, of the
target space.
Since the automorphism group is $U(1)$ (rather than $\R$), as a consequence of
the fermion
couplings, the two-form $F$ is an integral class in the cohomology of
spacetime, and
hence so is the pullback of $\Omega$. This imposes a condition on the target
manifolds
for which the model is consistent. We note here that this condition is clearly
satisfied by Hodge manifolds, for which the K\"ahler form is an integral class
in the
cohomology. This is very similar to the restriction on the target spaces of
D=3+1
sigma-models coupled to N=1 supergravity [\bagwit].
The fermion supersymmetry transformations in bosonic, on-shell, backgrounds
are
$$
\eqalign{
\delta\psi_m &= {\cal D}_m\epsilon\cr
\delta\lambda^A &= -{i\over2} f_I{}^A\gamma^m\epsilon\; \partial_m\phi^I }
\eqn\mfifteen
$$
Note that the Killing spinor condition $\delta\psi_m=0$ is unaffected by the
presence of
matter. However, its integrability condition \threeb\ is no longer
satisfied by all solutions of the field equations. In fact, using the field
equations
\mthirteen\ we find that the integrability condition becomes
$$
\big[\gamma^nT^m{}_n -j^m_{top}\big]\chi =0\ ,
\eqn\msixteen
$$
which can be rewritten as
$$
g^{IJ}\big(P_{IK}\gamma^n\partial_n\phi^K\big)\gamma^m
\big(P_{JL}\gamma^p\partial_p\phi^L\big)\chi =0
\eqn\mseventeen
$$
where $P$ is the projector
$$
P_{IJ} = g_{IJ} -i\Omega_{IJ}\ .
\eqn\meighteen
$$
By contracting \mseventeen\ with $\bar\chi$ we deduce that
$$
g^{IJ}\xi_I\gamma^m\xi_J =0
\eqn\mnineteen
$$
where
$$
\xi_I\equiv P_{IJ}\gamma^n\partial_n\phi^J\chi\ .
\eqn\mtwenty
$$
The left hand side of \mnineteen\ is manifestly timelike, and hence non-zero,
unless
$\xi_I=0$. Thus, the integrability condition for a Killing spinor in the
presence of
matter is equivalent to
$$
\xi_I=0 \ .
\eqn\mtwentyone
$$

For Killing spinors satisfying \threef, this condition is satisfied by
matter fields for which
$$
e_{\underline 0}{}^m\partial_m \phi^I =0\ ,\qquad\qquad
P_{IJ}\big( e_{\underline 1}{}^m \pm i e_{\underline 2}{}^m
\big)\partial_m\phi^J =0\ .
\eqn\mtwentytwo
$$
Choosing complex coordinates $\phi^\alpha$ on the target space, with complex
conjugates
$\bar\phi^{\bar\alpha}$, these equations become
$$
e_{\underline 0}{}^m\partial_m \phi^\alpha =0\ ,\qquad\qquad
e_{\pm}^m\partial_m\phi^\alpha\ ,
\eqn\mtwentythree
$$
where $e_{\pm}^m = e_{\underline 1}^m \pm ie_{\underline 2}^m$. For the static
spacetime metric
$$
ds^2 = dt^2 - e^{-\sigma(z,\bar z)} dz d\bar z\ ,
\eqn\mtwentyfour
$$
the first of eqs. \mtwentythree\ is trivially satisfied by time-independent
matter
fields while the second is satisfied by holomorphic functions, i.e
$$
\phi^\alpha = \phi^\alpha (z)\ .
\eqn\mtwentyfive
$$
The Einstein equation then implies that
$$
\partial\bar\partial \Big[\sigma(z,\bar z) - {\cal K}\big(\phi(z),\bar\phi(\bar
z)\big)\Big]=0
\eqn\mtwentysix
$$
where ${\cal K}(\phi,\bar\phi)$ is the K\"ahler potential of the target space
metric.
Similar equations have been considered previously, and solutions to them
discussed, in
the context of cosmic string solutions of 3+1 dimensional field theories
[\CG,\SCS].
We have now shown, in the context of the (2,0) Poincar\'e supergravity, that
solutions
of the form \mtwentyfour\ and \mtwentyfive\ satisfying \mtwentysix\ are
supersymmetric.
We expect that this is also true for the corresponding cosmic string solutions
in 3+1
dimensions.
\chapter{An energy theorem}
A customary feature of spacetimes admitting Killing spinors is that they
saturate a
Bogomol'nyi-type bound on the energy. To investigate this possibility here we
proceed
as usual by defining the Nester tensor
$$
E^{mn}=-{i\over 2}\bar \chi\gamma^{mnp}{\cal D}_p\chi +c.c. \ .
\eqn\threep
$$
Given a non-singular Cauchy surface $\Sigma$, we consider the integral
$\int_\Sigma dS_m
D_n E^{mn}$. Using the field equations, we obtain
$$
\int_\Sigma dS_m\, D_n E^{mn}=-\int_\Sigma dS_m \big[
{i\over 2}  {\overline {{\cal D}_n\chi}} \gamma^{mnp}{\cal
D}_p \chi+c.c.\big]\  +\ \int_\Sigma dS_m K^m\ ,
\eqn\tresc
$$
where
$$
K^m = -{i\over 2} G^m{}_n {\bar
\chi}\gamma^n\chi-ie^{-1}\varepsilon^{mnp}\partial_n A_p \bar \chi \chi \ .
\eqn\atresc
$$

Consider first the left hand side of \tresc. We may choose the spacelike
surface $\Sigma$
to be a surface of constant $t$, so that the only non-zero component of $dS_m$
is $dS_0$.
Then
$$
\int_\Sigma dS_m \,  D_n E^{mn}=\oint d\varphi\int
dr\,\big[ \partial_r ( \bar\chi{\cal D}_\varphi\chi)-\partial_\varphi(\bar\chi
{\cal D}_r\chi)\big]\ .
\eqn\trese
$$
The first term in the integral on the right hand side of this equation equals
$$
\oint_\infty d\varphi\, \bar\chi{\cal
D}_\varphi\chi=\oint_\infty d\varphi\, \big( \bar\chi \bar D_\varphi
\chi+{i\over 2}\bar\chi
(\omega^a_\varphi-\bar\omega^a_\varphi)\gamma_a\chi
-iA_\varphi^\infty\bar\chi \chi\big)         \ .
\eqn\tresf
$$
In order for this to be a combination of the mass and the charge, as defined in
\twog\
and \twoi, we require $\chi\sim \chi_\infty$ as $r\rightarrow\infty$, where
$\chi_\infty=
e^{if(\phi)}\chi_0$ with $\chi_0$ a non-zero {\it constant} spinor, and that
$\omega^a_\varphi\rightarrow 0$, for $a\neq {\underline 0}$.
The second term in the integral on the right hand
side of \trese\ is zero if it is  convergent. Convergence requires that $\int
dr\,
\bar\chi{\cal D}_r\chi<\infty$, which means that ${\cal D}_r\chi$ must go to
zero {\it
faster} than $1/r$ as $r\rightarrow\infty$.
These conditions on $\omega$ and $\chi$ are guaranteed if the metric is of the
form
$g_{mn}=g_{mn}^\infty+h_{mn}$, where $g_{mn}^\infty$
is the metric given in \twof, and $h_{mn}$ has the properties
$$
 \lim_{r\rightarrow \infty} h_{mn}=0\quad ,\quad
 \lim_{r\rightarrow \infty}(r h_{mn,p})=0\ .
\eqn\tresg
$$
Then, if we use the fact that $\bar \chi = i\chi^\dagger\gamma^\uzero$, and
take
$$
\chi^\dagger_\infty\chi_\infty=1\ ,
\qquad \gamma^\uzero\chi_\infty= \pm\chi_\infty\ ,
\eqn\atresg
$$
the right hand side of \tresf\ reduces to ${1\over 2}(M\pm 2Q) +
\oint_\infty\!\!
d\varphi\, \bar \chi_\infty \bar D_\varphi \chi_\infty$, and we therefore
deduce that
$$
\int_\Sigma dS_m \,  D_n E^{mn} = {1\over 2}(M\pm 2Q) + \oint_\infty\!\!
d\varphi\, \bar \chi_\infty \bar D_\varphi \chi_\infty\ .
\eqn\btresg
$$
Consider now the right hand side of \tresc. Provided that the vector $K$
is future-directed timelike or zero, the second integral is non-negative.
This condition is satisfied for the supersymmetric sigma-model matter of the
previous
section. This can be seen, following the discussion in [\IT] of the adS case,
by using
the field equations \mthirteen\ to rewrite $K^m$ as
$$
\eqalign{
K^m &=  -{i\over4} \bar\chi \big( T^m{}_n\gamma^n - j^m_{top}\big)\chi\cr
&= -{i\over4} g^{IJ}\xi_I\gamma^m\xi_j \ ,}
\eqn\newone
$$
where $\xi_I$ was defined in \mtwenty. The right hand side is a manifestly
future-directed timelike vector field, as required.
Thus, if we can establish that the first integral is also non-negative we will
have
established a bound on the mass $M$.  The first integral is indeed
non-negative if the spinor
$\chi$ is chosen to satisfy the 2+1 analogue of the Witten condition,
$$
{}^{(2)}g^{ij}\gamma_i{\cal D}_j\chi=0\ ,
\eqn\tresh
$$
where ${}^{(2)}g^{ij}$ is the inverse of the spatial part of the
metric. It turns out, however, that the only non-zero solutions of \tresh\ with
the
required boundary conditions are Killing spinors, so that instead of the
expected
inequality we find an equality. We now explain this point.
 From the asymptotic conditions \tresg\ it is easy to see that
$$
\eqalign{
 \lim_{r\rightarrow \infty}&r^2 g^{r\varphi}=0\quad ,\quad
  \lim_{r\rightarrow \infty}g^{rr}=-{1\over \lambda^2}\quad , \quad
  \lim_{r\rightarrow \infty}r^2 g^{\varphi\varphi}=-1\ ,\cr
   \lim_{r\rightarrow \infty}&\gamma_r=\lambda\gamma_\uuno\quad ,\quad
  \lim_{r\rightarrow \infty}{1\over r}\gamma_\varphi =\gamma_\udos\ ,\cr
  \lim_{r\rightarrow \infty}& {\cal
D}_\varphi\chi=\Big[\partial_\varphi\pm {1\over 2}\Big(1-{M\over
2\pi}\Big)-{Q\over 2\pi}\Big]\chi_\infty \quad, \quad
  \lim_{r\rightarrow \infty}r{\cal D}_r\chi=0\ .\cr}
\eqn\tresi
$$
Using this, it is clear that
$$
     0=\lim_{r\rightarrow \infty}r\ {}^{(2)}g^{ij}\gamma_i{\cal
D}_j\chi= -\gamma_\udos \Big[\partial_\varphi\pm {1\over 2}\Big(1-{M\over
2\pi}\Big)-{Q\over 2\pi}\Big]\chi_\infty\ .
\eqn\tresj
$$
This last equation tells us that $\chi_\infty$ is a Killing spinor of a conical
spacetime of charge $Q$ and mass $M$. We have shown earlier that such spinors
exist
only if either $M=2|Q|$ or $M=2|Q\pm \pi|$. Moreover, \tresj\ implies that the
right
hand side of \btresg\ vanishes which in turn implies that the spinor $\chi$ is
a
Killing spinor of the spacetime and not merely asymptotic to one. Thus, {\it
the
Witten-Nester method applied to 2+1 Poincar\'e gravity leads to an equality
rather
than a bound}!
In retrospect this result should not be so surprising because, in distinction
to
the adS case [\IT], the mass is determined by the {\it leading} terms in the
asymptotic
metric, i.e. by the boundary conditions. Note that this argument is independent
of the
presence of matter. Indeed, as noted in the previous section, the existence of
a
Killing spinor for a solution of the field equations requires that $\xi_I=0$,
which
implies that $K^m=0$.
We conclude this section with a comment on the role of angular momentum. A
spacetime
of the form [\DJtH]
$$
ds^2= \Big(dt+{J\over 2\pi}d\varphi\Big)^2-\lambda^2 dr^2+d\varphi^2 \ ,
\eqn\tresk
$$
where $J$ is the angular momentum, has exactly the same spin connection as
\twof, so the
Killing spinor equation is not altered. The argument leading to \tresj\ is not
modified
if $g^\infty_{mn}$ is taken to be \tresk\ so we arrive at the same conclusion
as
before. In other words, in contrast to the adS case, angular momentum is
irrelevant.
\chapter {Off-shell superfields}
As in four dimensions, it is convenient to discuss off-shell  superspace
supergravity
in three-dimensional spacetime starting from a superconformal perspective. One
possible definition of a superconformal structure on $N$-extended superspace
(the
$N=2$ case was discussed in [\vNRZ]) is as follows [\HL]: a superconformal
structure on
$(3|2N)$ dimensional superspace $M$ is a choice of odd tangent bundle $F$ (of
rank
$(0|2N)$) and a reduction of the structure group of the frame bundle of $F$ to
$G:=GL^+(2,R)\times_{Z_2} O(N)$ such that the Frobenius tensor of $F$ coincides
with
the natural tensor associated with $G$. The Frobenius tensor of the sub-bundle
$F$ of
the tangent bundle $T$ is defined by computing the commutators of vector fields
which
are sections of $F$ and evaluating them modulo $F$. This defines a tensor field
taking its values in $\Lambda^2 F^*\otimes B$ where $B=T/F$. The choice of $G$
means
that, locally at least, $F=S\otimes V$ where $S$ has rank $(0|2)$ and $V$ has
rank
$(N|0)$. The natural tensor associated with $G$ is the product of the metric on
$V$
with the Dirac matrices (considered as defining a map from from $\Lambda^2 S$
to $B$).
If we let $E_{\a i}$ denote a local basis of $F$ and $E^a$ a local basis of
$B^*$,
then these bases can be chosen such that the components of the Frobenius tensor
are
given by
$$
<[E_{\a i}, E_{\b j}], E^c>=\d_{ij} (\c^c)_{\a\b}
\eqn\foura
$$
where $<\ >$ denotes the pairing between vectors and forms.
In order to unravel the consequences of this structure it is  convenient to
make a
choice of $B$ as a subbundle of $T$ and to introduce a connection which we
choose to
take values in $sl(2,\R)\oplus o(N)$. These choices can be made in such a way
that
the components of the torsion tensor are, at dimension zero\foot{This
corresponds to a
rescaling of the generator $P_a$ in the supersymmetry algebra of \onea.},
$$
T_{\a i\b j}{}^c=-\d_{ij} (\c^c)_{\a\b}
\eqn\fourb
$$
(so that this part of the torsion is identified with the Frobenius tensor); at
dimension one-half,
$$
T_{\a i b}{}^c=T_{\a i\b j}{}^{\c k}=0\ ;
\eqn\fourc
$$
at dimension one,
$$
T_{a b}{}^c=0\ ;
\eqn\fourd
$$
and
$$
T_{a \b j}{}^{\c k} = i(\c^b)_{\b}{}^\c G_{ab j}{}^k -i (\c_a)_\b{}^\c\,
K_j{}^k
\eqn\foure
$$
while at dimension three-halves the leading component of the torsion in a
$\theta$-expansion is essentially the field strength tensor of the gravitini.
The tensor
$G$ is antisymmetric on both its internal and Lorentz indices while $K$ is
symmetric. The
components of the curvature tensor can be computed using the Bianchi
identities, and for
general $N$ one finds that the geometry is described by the two superfields $G$
and $K$
together with a third, which for $N\geq 4$ is a dimension one scalar
$A_{ijkl}$ totally antisymmetric on its internal indices. The component fields
corresponding to this structure divide into a conformal supergravity (CSG)
multiplet and
a compensating Weyl multiplet which is an entire scalar superfield. The CSG
multiplet is
$$
e_m{}^a\, ;\, \psi_{mi}\, ;\, A_{mij}, A_{ijkl}\, ;\, \r_{ijk}, \s_{ijklm}\,
;\,
B_{ijkl},
\dots
\eqn\fourf
$$
where each field is antisymmetric on its internal indices. The leading
components of
the fields $K$ and $G$ above belong to the Weyl multiplet. There are two
special
cases, $N=4$ and $N=8$. In $N=4$ one can reduce the internal symmetry group
from
$O(4)$ to $SO(3)$, in which case the CSG multiplet is simply
$(g_{mn},\psi_{mi},
A_{mij})$ where the gauge field $A_{mij}$ is now self-dual on its internal
indices. In
$N=8$ one can impose self-duality on the scalar field $A_{ijkl}$. The first few
components of a scalar superfield are
$$
A\, ;\, \l_i\, ;\, K_{ij},G_{a ij}\, ;\, \chi_{a ijk}\dots
\eqn\fourg
$$
where $G$ and $\chi$ are antisymmetric on their internal indices, $K$  is
symmetric
and $\chi$ is gamma-traceless.
We begin the discussion of Poincar\'e supergravity with $N=2$. The
compensating
multiplet can be constrained to be chiral, by taking $K_{ij}=\tilde K_{ij}$
(tracefree), or to be real linear by taking $K_{ij}=\d_{ij} K$. In the former
case the
field $G_a=\der_a B$, at least at the linearised level, while in the latter
case
$G_a$ is conserved. At the linearised level these multiplets are constructed
from a general real scalar superfield by
imposing the constraints
$$
D^2 A=0
\eqn\fourh
$$
or
$$
\tilde D^{ij}A=0
\eqn\fouri
$$
respectively, where $D^2$ and $\tilde D^{ij}$ are defined by
$$
D_{\a i} D_{\b j}=\e_{\a\b}\e_{ij} D^2 + \e_{\a\b} \tilde D^{ij} +
{1\over2} \d_{ij} (\c^a)_{\a\b}\der_a \ .
\eqn\fourj
$$
The chiral compensator corresponds to type (1,1) Poincar\'e
supergravity. The fields $A,B$ and $\l$ get absorbed by the metric,
gravitino and $SO(2)$ gauge field respectively to yield the following
multiplet:
$$
e_m{}^a\, ;\, \psi_{mi}\, ;\,  A_{m}, \tilde K_{ij}\ .
\eqn\fourk
$$

To implement this in superspace it is more convenient to reduce the  structure
group
to $SL(2,\R)$ and to combine the Weyl and $SO(2)$ parameters into a complex
scalar
superfield which can then be constrained to be chiral. The corresponding
constraints
resemble the $N=1$ minimal constraints in four dimensions and can be solved in
the
Ogievetsky-Sokatchev formalism [\OS]; alternatively, one can observe that the
action
is $\int_M E$ where $E$ is the superdeterminant of the
supervielbein, and derive the equations of motion by solving the constraints
for a
deformation of the supervielbein [\WZ]. The equations of motion imply that the
superspace is flat, as expected.
In the (2,0) case, with $K_{ij}=\d_{ij} K$, the Poincar\'e  supermultiplet that
is
obtained by combining the conformal SG multiplet with the compensator is
$$
e_m{}^a ,C_m; \psi_{mi}\, ;\,  A_m, K\ ,
\eqn\fourl
$$
where $C_m$ is a gauge field whose field strength is $G$ and $A_m$ is  the
$SO(2)$
gauge field. This theory resembles closely new minimal supergravity in four
dimensions [\SW] and is best described in superspace by introducing an abelian
gauge
field $C$ with field strength $G=d C$. The components of $G$ are, at dimension
zero,
$$
G_{\a i\b j}=-i\e_{\a\b}\e_{ij}\ ;
\eqn\fourm
$$
at dimension one-half,
$$
G_{a \b j}=0\ ,
\eqn\fourn
$$
while at dimension one we find $G_{ab}$ which is identified with the  field
occuring
in the torsion. Again the constraints can be solved in Ogievetsky-Sokatchev
fashion,
this time in a superspace with an extra bosonic coordinate to accommodate $C$.
Alternatively one can follow [\HST]. In this approach one observes that there
is a
choice of gauge in which the spinorial part of the $SO(2)$ gauge field takes
the form
$$
A_{\a +}=iD_{\a +} V\ ,
\eqn\fouro
$$
where $V$ is a real prepotential and $D_{\a +}={1\over\sqrt{2}}(D_{\a  1}-i
D_{\a
2})$. The action is $\int_M EV$. The corresponding component action differs
from the
CS one \twod\ only by the addition of a term $K^2$.
We now turn to the more interesting case of $N=3$. We first consider  (2,1)
supergravity. The internal part of the structure group is to be reduced from
$SO(3)$
to $SO(2)$, and so the compensating multiplet involves a total of three scalar
superfields corresponding to the Weyl and $O(3)/O(2)$ parameters. This triplet
of
fields can be constrained to form a vector multiplet, which has components
$$
A,A_i,C_m\, ;\,  \l,\l_i, \l'\, ;\,  K, K_i
\eqn\fourp
$$
where $i=1,2$. Combining this multiplet with the CSG multiplet
we find the following $(2,1)$ Poincar\'e supermultiplet
$$
e_m{}^a,C_m\, ;\, \psi_{m i}, \psi_m\l\, ;\,  K, K_i, A_m, A_{mi}\, ;\,  \r
\eqn\fourq
$$
where $A_m$ is the automorphism gauge field and the remaining dimension one
vectors are
non-gauge auxiliaries. We have not constructed the superspace action for this
theory,
but it should be possible to construct one using harmonic superspace techniques
[\KZ]. The off-shell component action differs from the CS action by the
addition of
terms quadratic in the bosonic auxiliaries $A_{mi},K,K_i$, and a term of the
form $\l\r$
which takes care of the auxiliary fermions.
The (3,0) theory is more complicated. In this case we have only the  Weyl
compensator
at our disposal. If we impose $K_{ij}=\d_{ij} K$, as in the (2,0) case but now
$i,j=1,2,3$, we find the following multiplet:
$$
A;\l_i\, ;\, G_{a i}\, ;\, \chi_a\ ,
\eqn\fourr
$$
where $G_{a i}$ is conserved and $\chi_a$ is gamma-traceless and
conserved. On the other hand the conformal multiplet is
$$
e_m{}^a\, ;\, \psi_{m i}\, ;\, A_{m i}\, ;\,  \r\ .
\eqn\fours
$$
The problem lies with the dimension three-halves fields $\r$ and  $\chi$. It
seems
plausible that these should be combined into a conserved, but not
gamma-traceless
field $\S_a$, so that, at the linearised level at least,
$$
\S_a= \e_{abc} \der^b \Lam^c\ .
\eqn\fourt
$$
The field $\Lam$ is a new gravitino. This suggests that the off-shell (3,0)
theory is
in fact an $N=4$ theory with component field content
$$
e_m{}^a, C_{m ij}\, ;\, \psi_{m i}\, ;\, A_{mij},K\ ,
\eqn\fouru
$$
where now $i,j=1,..,4$ and both $C$ and $A$ are self-dual on their $SO(4)$
indices.
This hypothesis can be tested by constructing an appropriate $N=4$
supergeometry, but
this is not entirely straightforward owing to the fact that the central charge
acts
non-trivially off-shell. However, it appears that this can be done, although we
have
neither verified it completely nor constructed a superspace action.
In outline, the construction starts with a superspace $M'$ of  dimension
$(3+3|8)$
which can be thought of as some sort of affine bundle over $N=4$ superspace
$M$. The
structure group of $M'$ geometry is taken to be $SL(2,\R)\xz SO(3)$. A
preferred set
of coframes is denoted by $E^{\uA}=(E^A, E^I)=(E^a, E^{\a i}, E^I)$, where
$i=1,..,4$
and $I=1,2,3$. The connection, $\C_{\uA}{}^{\uB}$ has non-vanishing components
$\C_a{}^b$, $\C_{\a i}{}^{\b j}=\d_\a{}^\b A_i{}^j+\d_i{}^j
\C_\a{}^\b$ and $\C_I{}^J$, where
$$
\C_{\a\b}=-{1\over4} (\c^{ab})_{\a\b} \C_{ab}
\eqn\fourv
$$
and
$$
\eqalign{
\C_{IJ}&=\e_{IJ}{}^K A_K \cr
A_{ij}&=f_{ij}{}^K A_K  \cr}
\eqn\fourw
$$
with $f_{ij}{}^K$ denoting the numerically invariant self-dual  tensor. The
non-vanishing components of the torsion are, at dimension zero,
$$
\eqalign{
T_{\a i\b j}{}^c&=-\d_{ij}(\c^c)_{\a\b} \cr
T_{\a i\b j}{}^K&=-i\e_{\a\b} f_{ij}{}^K \ ,\cr}
\eqn\fourx
$$
and at dimension one,
$$
\eqalign{
T_{a\b j}{}^{\c k} &= -i(\c_a)_\b{}^\c \d_j{}^k K +
{i\over2}(\c^b)_\b{}^\c G_{ab j}{}^k \cr T_{I\b j}{}^{\c k}
&={2\over3}\e_{\b\c} f_{jkI} K+{1\over4} f_{jk}{}^J\e_{IJK}
G_{\b\c}{}^K\cr T_{ab}{}^K &= G_{ab}{}^K \cr T_{IJ}{}^c &=
{1\over2}\e_{IJK} G^{c K} \ .\cr}
\eqn\foury
$$
This set of constraints is consistent with the Bianchi identities up  to
dimension
one, which is as far as we have checked. We are encouraged to believe that it
is
fully consistent as the dimension one identities are non-trivial.
For higher $N$ it seems to be more difficult to construct off-shell superspace
formalisms which can be used to write down actions. This is because the
multiplets
for general $N$ contain high spin, high dimension, fields and involve {\it
covariant}
conservation conditions. One might think that it would be possible to mimic the
CS
formalism in superspace, but again one runs into difficulties for higher $N$.
In this
approach one introduces a superspace gauge potential ${\cal A}$ which takes its
values in
the (p,q) Poincar\'e superalgebra. Because this includes supertranslations, the
correct
equations of motion are simply ${\cal F}(=d{\cal A}+{\cal A}^2)=0$. However,
because
of high spin component fields it is difficult to construct an action which
would
lead to this equation for all $N$. In fact this is true even for ordinary
supersymmetric Yang-Mills for which the conventional superspace approach works
only for
$N\leq 3$ [\NG] and the harmonic approach for $N$ up to 6 [\HLb]. It is not
ruled out
that superspace actions exist for all $N$, but, as yet, it is not clear how to
construct them.
We conclude this section with a brief discussion of matter coupling in (2,0)
supergravity
in the superspace formalism. To do this it is convenient to begin again with
$N=2$
superconformal geometry. This may be reformulated in a way which emphasises its
close
affinity with K\"{a}hler geometry as follows: one has a real supermanifold $M$
of dimension
$(3|4)$ with a choice of odd tangent bundle, $F$, such that $F$ is equipped
with a fibre
complex structure, $I\in\Gamma(End F),\ I^2=-1$, and such that the Frobenius
tensor,
regarded as a $B$-valued form in $\wedge^2 F^*$ is of type (1,1) with respect
to $I$. One
can then verify that such a supermanifold is in fact a $CR$ supermanifold,
which is to say
that, if $F_c={\cal F}\oplus\overline{\cal F}$, where $F_c$ is the
complexification of $F$,
then $\overline{\cal F}$ is involutive. This allows one to define a $CR$
exterior
derivative, $\bar D$, satisfying $\bar D^2=0$.
The structure group associated with the above superconformal structure is
$SL(2,R)\cdot
U(1)\times R^*$; it can be reduced to $SL(2,R)\cdot U(1)$ by the introduction
of a fibre
metric $g_{F}\in\Gamma(S^2 F^*)$, which can be taken to be hermitian. This then
allows one
to define a fermionic K\"{a}hler 2-form $\omega_{F}\in\Gamma(\wedge^2 F*)$ by
lowering the
contravariant index on $I$ using the fibre metric. The two-form $\omega_F$ is
in fact a
$(1,1)$ form and is furthermore $\bar D$-closed. It extends to a closed
two-form on the
whole space if and only if the superfield $K_{ij}=\delta_{ij}K$. The two-form
obtained in
this way is precisely the two-form $G$ introduced earlier.
The matter field $\phi$ is introduced as a $CR$ map $\phi: M\rightarrow N$,
where $N$ is
the target space which is taken to be K\"{a}hler, that is $\phi$ is complex and
chiral,
$\bar D\phi=0$. The action for (2,0) supergravity coupled to matter is then
given by
$$
S=\int_M E(V-{\cal K})
\eqn\fourz
$$
where ${\cal K}$ is the K\"{a}hler potential of $N$ and $V$ is the prepotential
for the
$U(1)$  gauge field introduced earlier. The equation of motion for $\phi$ is
$$
\nabla_{\alpha +} D^{\alpha}_+ \phi^r=0;\qquad r=1,\ldots {\rm dim_c\ N},
\eqn\fourzone
$$
where $\nabla$ includes the (pull-back of) the Levi-Civita connection on $N$.
The
equations of motion resulting from the variation of the supergravity fields can
be
expressed in the following simple form:
$$
F=\phi^*\Omega,
\eqn\fourztwo
$$
where $F$ is the $U(1)$ field strength two-form and $\Omega$ the K\"{a}hler
two-form  for
$N$. These equations are completely equivalent to the component results
obtained in section
4 as one may easily verify. One observes again the requirement that the target
manifold
should be Hodge as a consequence of the equations of motion.
\chapter{(p,q) Poincar\'e theories}
The CS formulation of the general (p,q) Poincar\'e supergravity theory is found
as
follows. We first divide the N spinor supercharges of the MS supergravity
theory
into a set of $p$ charges $Q^i,\, i=1,\dots, p$, and the complementary set of
$q$
charges $Q^{i'},\, i'=1,\dots, q$, and introduce $p(p-1)/2 +q(q-1)/2$ central
charges,
$Z^{ij}=-Z^{ji}$ and $Z^{i'j'}=-Z^{j'i'}$. We then define the
(p,q)-Poincar\'e superalgebra to be the one for which the non-trivial
(anti)commutators
are
$$
\eqalign{
\{Q_\alpha^i,Q_\beta^j\} &= -{1\over2}\delta^{ij}(\gamma^a)_{\alpha\beta} P_a +
i
\epsilon_{\alpha\beta}Z^{ij}\cr
\{Q_\alpha^{i'},Q_\beta^{j'}\} &=
-{1\over2}\delta^{i'j'}(\gamma^a)_{\alpha\beta}P_a -
i\epsilon_{\alpha\beta}Z^{i'j'} \cr
[M_a,M_b] &= -\epsilon_{ab}{}^{c}M_c \cr
[M_a,P_b] &= -\epsilon_{ab}{}^{c}P_c \cr
[M_a,Q_\alpha^i] &= {i\over2}(Q^i\gamma_a )_\alpha \cr
[M_a,Q_\alpha^{i'}] &= {i\over2}(Q^{i'}\gamma_a)_\alpha  \ .}
\eqn\fiveaa
$$
When $p$ or $q$ is greater than unity, the (p,q) Poincar\'e superalgebra is
a central extension of the N-extended Poincar\'e superalgebra. In this case is
not
possible to formulate a CS action for the algebra \fiveaa\ because it does not
admit a
non-degenerate invariant inner product. To see this we note first that the
inner
product $\la\; ,\;\ra$ of an even generator with an odd generator must vanish.
Then,
since $\la even,odd\ra$ is both zero and invariant we deduce, in particular,
that
$$
-{1\over2}\delta^{ij} (\gamma^a)_{\alpha\beta}\la P_a,Z^{kl}\ra
+\epsilon_{\alpha\beta}\la Z^{ij}, Z^{kl}\ra =0
\eqn\fiveab
$$
which implies that both $\la P,Z\ra$ and $\la Z, Z\ra$ vanish. Furthermore, the
invariance of $\la M,Z\ra$ implies immediately that $\la M,Z\ra=0$. Similar
arguments
apply to the
$Z'$ generators so the only remaining way to achieve a non-degenerate invariant
inner
product would be to require $p=q$ and to pair the $Z$ with the $Z'$ generators.
But
starting from the invariance of $\la Q',Z\ra$, a similar argument to the above
one
shows that
$\la Z,Z'\ra=0$. Thus, $Z$ and $Z'$ are orthogonal to all generators of the
(p,q)
Poincar\'e superalgebra, including themselves, so any invariant inner product
is
degenerate.
For this reason, we consider an enlarged superalgebra obtained in the following
way.
We first observe that the (p,q) Poincar\'e superalgebra has an $SO(p)\times
SO(q)$
automorphism group\foot{The 'automorphism group' of a centrally extended
Poincar\'e superalgebra has been often defined in the supersymmetry literature
as that
subgroup of the automorphism group of the algebra without central charges that
commutes
with the central charges. Here we adopt the standard mathematical terminology
in which
the central charges of an algebra need not be invariant under that algebra's
automorphism group.}. We now take the semi-direct extension of this
$so(p)\oplus so(q)$ automorphism algebra by the (p,q)-Poincar\'e superalgebra.
Thus, we
now include the new generators, $T^{ij}=-T^{ji}$, and $T^{i'j'}=-T^{j'i'}$,
with the new
non-trivial commutators as follows:
$$
\eqalign{
[ T^{ij}, Q^k] &= 2\delta^{k[j}Q^{i]}\cr
[ T^{i'j'}, Q^{k'}] &= 2\delta^{k'[j'}Q'{}^{i']}\cr
[ T^{ij}, Z^{kl}] &= 2[\delta^{k[j}Z^{i]l}- (k\leftrightarrow l)] \cr
[ T^{i'j'}, Z^{k'l'}] &= 2[\delta^{k'[j'}Z^{i']l'}- (k'\leftrightarrow l')] \cr
[ T^{ij}, T^{kl}] &= 2[\delta^{k[j}T^{i]l}- (k\leftrightarrow l)]\cr
[ T^{i'j'}, T^{k'l'}] &= 2[\delta^{k'[j'}T^{i']l'}- (k'\leftrightarrow l')] }
\eqn\fivea
$$
Note that although the charges $Z$ and $Z'$ are central in \fiveaa\ they are no
longer
central in the extended algebra including \fivea. The new superalgebra admits
an
invariant non-degenerate inner product, for which the non-vanishing components
are
$$
\eqalign{
\la M_a,P_b\ra &= \mu\eta_{ab}\cr
\la M_a,M_b\ra &= \lambda\eta_{ab}\cr
\la Q_\alpha^i,Q_\beta^j\ra &=i \mu\varepsilon_{\alpha\beta}\delta^{ij}\cr
\la Q_\alpha^{i'},Q_\beta^{j'}\ra &=
i\mu\varepsilon_{\alpha\beta}\delta^{i'j'}\cr
\la Z^{ij},T^{kl}\ra &= -2\mu\delta^{k[i}\delta^{j]l} \cr
\la Z^{i'j'},T^{k'l'}\ra &= 2\mu\delta^{k'[i'}\delta^{j']l'} \cr
\la T^{ij}, T^{kl}\ra &=2\xi\delta^{k[i}\delta^{j]l} \cr
\la T^{i'j'}, T^{k'l'}\ra &=2\xi'\delta^{k'[i'}\delta^{j']l'} \ ,}
\eqn\fiveb
$$
where $\mu$ is a non-zero real constant and, $\lambda$, $\xi$ and $\xi'$ are
arbitrary
real constants.
We introduce the gauge field one-forms associated with the generators of
this enlarged superalgebra via the connection one-form
$$
a =\omega^aM_a + e^aP_a + {1\over2} C_{ij}Z^{ij} +{1\over2}
C'_{i'j'}Z'{}^{i'j'}
+ {1\over2}A^{ij}T_{ij} + {1\over2}A'{}^{i'j'}T'{}_{i'j'}
+ \psi^iQ_i + \psi^{i'} Q_{i'}\ ,
\eqn\fivec
$$
where the coefficients of the generators are the one-form gauge potentials.
The corresponding field strength $F=da +a^2$ is
$$
\eqalign{
F= T^a P_a + &F^a(\omega)M_a + {1\over2}F^{ij}(A)T_{ij} +
{1\over2}F^{i'j'}(A')T_{i'j'}
\cr &
+ {1\over2}G^{ij}(C)Z_{ij}
+ {1\over2}G^{i'j'}(C')Z_{i'j'} + [{\cal D}\psi^i] Q^i + [{\cal D'}\psi^{i'}]
Q^{i'} \ ,}
\eqn\fived
$$
where
$$
\eqalign{
T^a &= de^a-\epsilon^a{}_{bc}\omega^b e^c -{1\over 4}{\bar
\psi}^i\gamma^a \psi^i-{1\over 4}{\bar \psi}^{i'}\gamma^a\psi^{i'}\cr
F^a(\omega) &=d\omega^a -{1\over2}\epsilon^a{}_{bc}\omega^b\omega^c\cr
F^{ij}(A) &= dA^{ij}+ A^{ik}A^{kj}\cr
F^{i'j'}(A') &= dA^{i'j'}+ A^{i'k'}A^{k'j'} \cr
G^{ij}(C) &= dC^{ij}+ { C}^{ik}A^{kj}+A^{ik}{C}^{kj}-i\bar\psi^i\psi^j \cr
G^{i'j'}(C') &= dC^{i'j'}+  C^{i'k'}A^{k'j'}+A^{i'k'}
C^{k'j'}+i\bar\psi^{i'}\psi^{j'}\cr
{\cal D}\psi^i&= d\psi^i+{i\over 2}\omega^c\gamma_c\psi^i+A^{ij}\psi^j\cr
{\cal D'}\psi^{i'} &=
d\psi^{i'}+{i\over2}\omega^c\gamma_c\psi^{i'}+A^{i'j'}\psi^{j'}\ .}
\eqn\fivee
$$
The CS action can be written, up to a surface term, as
$$
\eqalign{
S=\int\!d^3x\Big[ &2\mu e_aF^a(\omega) + \lambda Q_3(\omega) - i
\mu \bar\psi^i {\cal D}\psi^i - i \mu \bar\psi^{i'} {\cal D}\psi^{i'}
-{\xi\over2}Q_3(A) -{\xi'\over2}Q_3(A')\cr
& +\mu\, C^{ji}F^{ij}(A) -\mu\, C^{j'i'}F^{i'j'}(A') \Big]\ , }
\eqn\fivef
$$
where
$$
\eqalign{
Q_3(\omega) &=  \omega_a d\omega^a -
{1\over3}\epsilon_{abc}\omega^a\omega^b\omega^c\cr
Q_3(A) &= A^{ij}dA^{ji} + {2\over3} A^{ik}A^{kj}A^{ji}\cr
Q_3(A') &= A^{i'j'}dA^{j'i'} + {2\over3} A^{i'k'}A^{k'j'}A^{j'i'}  \ .}
\eqn\fiveg
$$
This action is invariant up to a surface term under the gauge transformation of
the connection $a$. In particular, the non-zero supersymmetry transformation
laws of the
fields are
$$
\eqalign{
\delta e^a&={1\over2} \bar\zeta^i\gamma^a\psi^i+{1\over2}
\bar\zeta^{i'}\gamma^a\psi^{i'}
\cr
\delta C^{ij}&=- 2i \bar \psi^{[i}\zeta^{j]}
\cr
\delta C^{i'j'}&=2i \bar \psi^{[i'}\zeta^{j']}
\cr
\delta \psi^i &= {\cal D}\zeta^i \cr
\delta \psi^{i'} &= {\cal D}'\zeta^{i'}\ ,}
\eqn\fiveh
$$
where $\zeta$ and $\zeta'$ are the anticommuting spinor parameters. Observe
that
the automorphism gauge fields enter these transformations. It is this fact that
accounts for the existence of Killing spinors for asymptotically conical
spacetimes.
This can be seen by repeating the analysis of section 3. One finds that Killing
spinors
exist provided that the holonomy of the $SO(p)\times SO(q)$ connections is
reduced to a
product of $U(1)$ factors.
Note that the parameters $\lambda,\xi$ and $\xi'$ appearing in the action
\fivef\ have
dimension $-1$ in mass units, relative to $\mu$. It follows that the Lagrangian
has a
definite scaling weight only if $\lambda=\xi=\xi'=0$. In this case \fivef\
simplifies to
$$
S=\mu \int\!d^3x\Big[ 2e_aF^a(\omega)  - i
\bar\psi^i {\cal D}\psi^i - i \bar\psi^{i'} {\cal D}\psi^{i'}
+ C^{ji}F^{ij}(A) - C^{j'i'}F^{i'j'}(A') \Big]\ ,
\eqn\fivef
$$
which reduces to the action \twod\ for the (2,0) case.
We remark that the above construction also works for `non-standard'
supersymmetry algebras with non-compact automorphism groups, obtained by
replacing the
Kronecker deltas $\delta^{ij}$, $\delta^{i'j'}$  by the invariant tensors of
non-compact versions of $SO(p)\times SO(q)$. In the (3,0) case such a
non-standard
supergravity theory, with automorphism group $SL(2;\R)$ instead of $SO(3)$, has
the advantage that the holonomy matching condition for Killing spinors can be
satisfied by cancelling the holonomy of the gravity sector against the holonomy
of the
gauge connection.
\chapter{adS supergravity and the Poincar\'e limit}
We now turn to the relation of the (p,q) Poincar\'e supergravity theories with
the
(p,q) adS theories. Unlike the Poincar\'e superalgebra, the adS superalgebras
are
semi-simple so the algebra of the outer automorphisms is isomorphic to the
algebra of
the inner automorphisms. It follows that the extended superalgebra that
includes
the automorphism generators is necessarily isomorphic to a direct
sum of the adS superalgebra and its automorphism algebra. Consequently, nothing
essential is gained in the formulation of the adS theories by the inclusion
of automorphism generators. However, since the new (p,q) Poincar\'e
supergravity
theories include additional $SO(p)\times SO(q)$ gauge fields, it is clear that
to
obtain these Poincar\'e theories as limits of adS theories we must enlarge the
adS superalgebras accordingly.
We begin with the direct sum of the standard (p,q) adS superalgebra and an
$so(p)\oplus so(q)$ algebra. The non-zero (anti)commutation relations are
$$
\eqalign{
[M_a,M_b] &= -\epsilon_{ab}{}^cM_c\cr
[P_a,P_b] &= -4m^2\epsilon_{ab}{}^cM_c\cr
[M_a,P_b] &= -\epsilon_{ab}{}^cP_c\cr
\{ Q^i_\alpha,Q^j_\beta\} &=-{1\over2}\big(\gamma^a\big)_{\alpha\beta}P_a
\delta^{ij} -
m \big(\gamma^a\big)_{\alpha\beta}M_a \delta^{ij} +i
m\epsilon_{\alpha\beta}Z^{ij}\cr
\{ Q^{i'}_\alpha,Q^{j'}_\beta\} &=-{1\over2}\big(\gamma^a\big)_{\alpha\beta}P_a
\delta^{i'j'} +m \big(\gamma^a\big)_{\alpha\beta}M_a \delta^{i'j'}
-i m\epsilon_{\alpha\beta}Z^{i'j'}\cr
[M_a,Q^i_\alpha ] &= {i\over2}\big( Q^i\gamma_a\big)_\alpha\cr
[P_a,Q^i_\alpha ] &= im\big( Q^i\gamma_a\big)_\alpha\cr
[M_a,Q^{i'}_\alpha ] &= {i\over2}\big( Q^{i'}\gamma_a\big)_\alpha\cr
[P_a,Q^{i'}_\alpha ] &= -im\big( Q^{i'}\gamma_a\big)_\alpha\cr
[Z^{ij}, Q^k_\alpha] &= 2\delta^{k[j}Q^{i]}_\alpha\cr
[Z^{i'j'}, Q^{k'}_\alpha] &= 2\delta^{k'[j'}Q^{i']}_\alpha\cr
[Z^{ij},Z^{kl}] &= 2[\delta^{k[j} Z^{i]l} - (k\leftrightarrow l)]\cr
[Z^{i'j'},Z^{k'l'}] &= 2[\delta^{k'[j'} Z^{i']l'} - (k'\leftrightarrow l')]\cr
[\bar T^{ij},\bar T^{kl}] &= -2 m[\delta^{k[j}\bar T^{i]l} - (k\leftrightarrow
l)]\cr
[\bar T^{i'j'},\bar T^{k'l'}] &=- 2m[\delta^{k'[j'}\bar T^{i']l'} -
(k'\leftrightarrow
l')]\ .}
\eqn\sixa
$$
The mass parameter $m$ determines the scale of the cosmological constant. We
now introduce the connection one-form
$$
\hat a =\omega^aM_a + e^aP_a + {1\over2} A_{ij}Z^{ij} +{1\over2}
A'_{i'j'}Z'{}^{i'j'}
+ {1\over2}\bar C^{ij}\bar T_{ij} + {1\over2}\bar C'{}^{i'j'}\bar T'{}_{i'j'}
+ \psi^iQ_i + \psi^{i'} Q_{i'}\ .
\eqn\athreea
$$
The curvature two-form $\hat F = d\hat a + \hat a^2$ is
$$
\eqalign{
\hat F =& T^a P_a + \hat F^a(\omega)M_a + {1\over2}\hat F^{ij}(A)Z_{ij} +
{1\over2}\hat  F^{i'j'}(A')Z_{i'j'} \cr &+ {1\over2}\bar G^{ij}(\bar C)\bar
T_{ij}
+{1\over2}\bar G^{i'j'}(\bar C')\bar T_{i'j'} + \Psi^i Q^i + \Psi^{i'} Q^{i'}\
, }
\eqn\bthreea
$$
where
$$
\eqalign{
\hat F^a(\omega) &= F^a(\omega) -2m^2
\epsilon_{abc} e^b e^c-{m\over 2}{\bar
\psi}^i\gamma_a\psi^i+{m\over 2}{\bar \psi}^{i'}\gamma_a\psi^{i'}\cr
\hat F^{ij} &= F^{ij}(A)-im{\bar \psi}^i\psi^j \cr
\hat F^{i'j'} &= F^{i'j'}(A') +im{\bar \psi}^{i'}\psi^{i'}\cr
\Psi^i &= {\cal D}\psi^i +ime^a\gamma_a\psi^i \cr
\Psi^{i'} &= {\cal D'}\psi^{i'} -ime^a\gamma_a\psi^{i'} \cr
\bar G^{ij} &= d\bar C^{ij}-m\bar C^{ik}\bar C^{kj}\cr
\bar G^{i'j'} &= d\bar C^{i'j'}-m\bar C^{i'k'}\bar C^{k'j'} \ ,}
\eqn\athreea
$$
and $T^a$, $F(\omega)$, $F(A)$, $F(A')$, ${\cal D}$ and ${\cal D}'$ are as
defined in
the previous section
The algebra \sixa\ has a class of invariant non-degenerate inner products
depending on at least three and at most four parameters. This inner product is
$$
\eqalign{
\la M_a,M_b\ra&=\lambda \eta_{ab}\ , \qquad  \la P_a,M_b\ra=\mu \eta_{ab}\ ,
\qquad \la P_a,P_b\ra=4 m^2\lambda \eta_{ab}\ ,
\cr
\la Q^i_\alpha, Q_\beta^j\ra&=i (\mu+2\lambda m) \epsilon_{\alpha \beta}
\delta^{ij}\ ,
\la Q^{i'}_\alpha, Q_\beta^{j'}\ra= i(\mu-2\lambda m) \epsilon_{\alpha \beta}
\delta^{i'j'}\ ,
\cr
\la Z^{ij}, Z^{kl}\ra&= 2(2\lambda+{\mu\over m}) ( \delta^{i[l} \delta^{k]j})\
,
\qquad \la Z^{i'j'}, Z^{k'l'}\ra= 2(2\lambda-{\mu\over m}) ( \delta^{i'[l'}
\delta^{k']j'})\ ,
\cr
\la \bar T^{ij}, \bar T^{kl}\ra&= 2\rho ( \delta^{i[l} \delta^{k]j})\ ,
\qquad \la \bar T^{i'j'}, \bar T^{k'l'}\ra= 2\rho' ( \delta^{i'[l'}
\delta^{k']j'})\ ,  }
\eqn\sixb
$$
where $\mu$, $\lambda$, and (for $p>1$) $\rho$ and (for $q>1$) $\rho'$ are free
real
parameters.
If one sets $m=0$ in \sixa\ the resulting algebra is the semi-direct
extension of $so(p)\oplus so(q)$ by the direct sum of the N-extended
Poincar\'e superalgebra with an abelian algebra of dimension
$\big[{p(p-1)\over2}+{q(q-1)\over2}\big]$. This contracted algebra is not
isomorphic to
the one given in section 2. However, it is known that the Wigner contractions
of
isomorphic algebras are not necessarily  isomorphic [\AA] ; indeed, the algebra
of
section 5, defined by \fiveaa\ and \fivea, results from a different contraction
of
\sixa. To see this, we use the redefinition
$$
T^{ij}=Z^{ij}-{1\over m}\bar T^{ij}\ , \qquad T^{i'j'}=Z^{i'j'}-{1\over m}\bar
T^{i'j'}\ ,
\eqn\sixd
$$
to eliminate $Z$ and $Z'$ in favour of $T$ and $T'$.  The non-trivial
commutators
involving the $T$ and $T'$ generators are then
$$
\eqalign{
\{ Q^i_\alpha,Q^j_\beta\} &=-{1\over2}\big(\gamma^a\big)_{\alpha\beta}P_a
\delta^{ij}
- m \big(\gamma^a\big)_{\alpha\beta}M_a \delta^{ij} + i
\epsilon_{\alpha\beta}(\bar
T+m T)^{ij}\cr
\{ Q^{i'}_\alpha,Q^{j'}_\beta\} &=-{1\over2}\big(\gamma^a\big)_{\alpha\beta}P_a
\delta^{i'j'} + m \big(\gamma^a\big)_{\alpha\beta}M_a \delta^{i'j'}
-i \epsilon_{\alpha\beta}(\bar T+m T)^{i'j'}\cr
[T^{ij}, Q^k_\alpha] &= 2\delta^{k[j}Q^{i]}_\alpha\cr
[T^{i'j'}, Q^{k'}_\alpha] &= 2\delta^{k'[j'}Q^{i']}_\alpha\cr
[T^{ij},T^{kl}] &= 2[\delta^{k[j} T^{i]l} - (k\leftrightarrow l)]\cr
[T^{i'j'},T^{k'l'}] &= 2[\delta^{k'[j'} T^{i']l'} - (k'\leftrightarrow l')]\cr
[T^{ij},\bar T^{kl}] &= 2 [\delta^{k[j}\bar T^{i]l} - (k\leftrightarrow l)]\cr
[T^{i'j'},\bar T^{k'l'}] &= 2 [\delta^{k'[j'}\bar T^{i']l'} -
(k'\leftrightarrow
l')]\ .}
\eqn\sixe
$$
Similarly the components of the inner product involving $T$ and $T'$ are
$$
\eqalign{
\la T^{ij}, T^{kl}\ra&= 2({\rho\over m^2}+2\lambda+{\mu\over m}) ( \delta^{i[l}
\delta^{k]j})\cr
\la T^{i'j'}, T^{k'l'}\ra &= 2({\rho'\over m^2}+2\lambda-{\mu\over m}) (
\delta^{i'[l'}
\delta^{k']j'})\cr
 \la T^{ij}, \bar T^{kl}\ra&=-{2\rho\over m}( \delta^{i[l} \delta^{k]j})\cr
\la T^{i'j'}, \bar T^{k'l'}\ra&=-{2\rho'\over m} ( \delta^{i'[l'}
\delta^{k']j'})\ . }
\eqn\sixf
$$
The $m\rightarrow 0$ limit of this algebra is the one of section 5. Moreover,
the inner
product used in section 5 for $\xi=-2\lambda$ and $\xi'=-2\lambda$ can be
obtained from
the inner product defined above by setting $\rho=-m\mu $ and $\rho{'}=m\mu$,
and then
taking the $m\rightarrow 0$ limit.
The CS action of (p,q) adS supergravity based on the algebra \sixa\ and the
inner product \sixb\ is, up to surface terms,
$$
\eqalign{
S=&\int\!d^3x\Big[ 2\mu e_aF^a(\omega) + \lambda Q_3(\omega) -
{4m^2\mu\over3}\epsilon_{abc}e^ae^be^c + 4m^2\lambda e_aDe^a\cr
&- i (\mu + 2\lambda m)[\bar\psi^i{\cal D}\psi^i +im\bar\psi^i\gamma_a
e^a\psi^i]
- i (\mu - 2\lambda m)[\bar\psi^{i'}{\cal D'}\psi^{i'} -im\bar\psi^{i'}\gamma_a
e^a\psi^{i'}] \cr
& +{1\over2}(2\lambda + {\mu\over m}) Q_3(A) +{1\over2}(2\lambda - {\mu\over
m})
Q_3(A') + {\rho\over2}Q_3(\bar C) + {\rho'\over2}Q_3(\bar C')\Big]}
\eqn\sixg
$$
where
$$
\eqalign{
Q_3(\bar C) &=  \bar C d\bar C - {2\over3} m\bar C^3 \cr
Q_3(\bar C') &= \bar C' d\bar C' - {2\over3} m\bar C'{}^3 \cr
De^a & =de^a-\epsilon^a{}_{bc} \omega^b e^c \ .}
\eqn\sixh
$$
To recover (p,q) Poincar\'e supergravity in the $m\rightarrow 0$ limit we set
$$
\bar C = C-{1\over m}A \qquad \bar C' =C'-{1\over m}A'
\eqn\sixi
$$
and choose
$$
\rho=-m\mu \qquad \rho' = m\mu\ .
\eqn\sixj
$$
In the $m\rightarrow 0$ limit we recover the action \fivef\ with the parameters
$\xi$
and $\xi'$ equal to $-2\lambda$.
\chapter{Comments}
We have constructed a new class of 2+1 Poincar\'e supergravity theories with
(p,q) supersymmetry by including additional gauge fields associated with
central charge
and automorphism generators. In contrast to previous N=(p+q) extended
Poincar\'e
supergravity theories, the new ones arise naturally as limits of an
$SO(p)\times SO(q)$
trivial extension of the (p,q)-supersymmetric adS supergravity theories. In
addition,
both the (1,1) and the (2,0) Poincar\'e supergravity theories have an off-shell
superfield formulation. The new (2,0) theory is analogous to the new-minimal
formulation of 3+1 supergravity. We have constructed the general coupling of
this theory
to sigma-model matter. Like the coupling to D=4 N=1 supergravity, the target
space is
required to be a Hodge manifold.
 From the superfield point of view, the existence of
distinct $N$-extended pure Poincar\'e supergravity theories with the same value
of $N$ can
be seen to be a consequence of the different possible choices of conformal
compensating
superfield. In 2+1 dimensions this choice can lead to inequivalent theories as
a result
of the Chern-Simons structure of pure 2+1 supergravity theories. For this
reason we do not
expect a similar phenomenon for conformal supergravity theories. One feature of
the new
Poincar\'e theories, e.g. (2,0), is that the conical spacetimes of charged
point
particles admit Killing spinors for special values of the masses and charges;
in this
sense these spacetimes are the 2+1 analogues of the Papapetrou-Majumdar multi
charged
black hole solutions of 3+1 Maxwell/Einstein theory.
The subdivision of the $N$-extended supergravity theories into (p,q) ones is
inevitable in the adS case because of the structure of the adS superalgebra.
For the
reasons just explained, it is also natural in the Poincar\'e case, although
here one can
envisage a more general subdivision into partitions $(p_1,p_2,\dots, p_k)$ with
$N=\sum_{i=1}^k p_i$. Such models can indeed be constructed, and the vacuum
spacetimes
again admit Killing spinors under suitable conditions.
Finally, it is of obvious interest to quantize the new supergravity theories.
The
additional gauge fields will provide an additional finite number of degrees of
freedom equal to the dimension of the moduli space of flat connections.
\vskip 1cm
\centerline{\bf Acknowledgements}
\vskip 0.5cm
G.P. is supported by a University Research Fellowship from the Royal Society.
J.M.I.
thanks the Commission of the European Community and CICYT (Spain) for financial
support.
\refout
\end